\documentclass[12pt]{article}
\usepackage{amsmath}
\usepackage{graphics}
\usepackage{centernot}

\usepackage{natbib}
\setcitestyle{authoryear,open={(},close={)}}

\usepackage[utf8]{inputenc}
\usepackage[T1]{fontenc}
\usepackage{babel}
\usepackage{rotating}

\usepackage{lscape}

\newcommand{\beginsupplement}{%
        \setcounter{table}{0}
        \renewcommand{\thetable}{S\arabic{table}}%
        \setcounter{figure}{0}
        \renewcommand{\thefigure}{S\arabic{figure}}%
               \setcounter{section}{0}
        \renewcommand{\thesection}{S\arabic{section}}%
     }

\newcommand{\bmx}{\mbox{\boldmath $x$}}
\newcommand{\bmX}{\mbox{\boldmath $X$}}
\newcommand{\bg}{\mbox{\boldmath $\gamma$}}
\newcommand{\bb}{\mbox{\boldmath $\beta$}}

\newcommand{\bn}{\mbox{\boldmath $\eta$}}

\newcommand{\bmZ}{\mbox{\boldmath $Z$}}
\newcommand{\bmz}{\mbox{\boldmath $z$}}

\newcommand{\CI}{\mathrel{\perp\mspace{-10mu}\perp}}

\def\stackunder#1#2{\mathrel{\mathop{#2}\limits_{#1}}}
 
\DeclareMathOperator*{\argmax}{arg\,max}

\title{Measuring the Impact of New Risk Factors Within Survival Models}
\author{Glenn Heller and Sean M. Devlin}
\date{%
    Memorial Sloan Kettering Cancer Center, New York, USA\\[2ex]%
}

\begin{document}
\maketitle

\begin{abstract}

Survival is poor for patients with metastatic cancer, and it is vital to examine new biomarkers that can improve patient prognostication and identify those who would benefit from more aggressive therapy. In metastatic prostate cancer, two new assays have become available: one that quantifies the number of cancer cells circulating in the peripheral blood, and the other a marker of the aggressiveness of the disease. It is critical to determine the magnitude of the effect of these biomarkers on the discrimination of a model-based risk score. To do so, most analysts frequently consider the discrimination of two separate survival models: one that includes both the new and standard factors and a second that includes the standard factors alone. However, this analysis is ultimately incorrect for many of the scale-transformation models ubiquitous in survival, as the reduced model is misspecified if the full model is specified correctly. To circumvent this issue, we developed a projection-based approach to estimate the impact of the two prostate cancer biomarkers. The results indicate that the new biomarkers can influence model discrimination and justify their inclusion in the risk model; however, the hunt remains for an applicable model to risk-stratify patients with metastatic prostate cancer.
\end{abstract}

\bigskip


\setcounter{page}{1}

\section{Introduction}  

Survival risk models are used to assess the risk of disease onset for an individual in the general population or the risk of disease progression or death for a patient already diagnosed with an illness.  The search to identify new risk factors and biomarkers to improve survival risk models is ubiquitous in clinical research.  For example in oncology, a large research effort is underway to evaluate patient risk as a function of molecular alterations within the tumor cell as assessed via a biopsy or peripheral blood.  An important research question is whether this newly acquired cellular information impacts survival risk, after accounting for existing clinical measures and laboratory tests obtained during routine practice.

In this work, the interest is focused on the added value of two new blood-based biomarkers, circulating tumor cells (CTC) and serum testosterone, in men with metastatic prostate cancer. Contemporary assays, such as CTC and serum testosterone, are typically more expensive to administer than the conventional assays used to assess patient risk in this population. To justify their routine use within a risk model, evidence beyond association metrics with survival is required. In this paper, we seek a second level of evidence, specifically examining how the incorporation of these factors into the risk model influences the concordance probability. 

Data from a randomized clinical trial \citep{Saad2015} of men treated on one trial arm with an androgen receptor targeted therapy, a key molecular target in the metastatic prostate cancer population, was used to evaluate the added value of CTC and serum testosterone. The primary endpoint of the clinical trial was overall survival, and the median survival time was only 31 months. This relatively short survival accentuates that accurate baseline risk determination is critical in the patient population. Historically, the prostate specific antigen (PSA) biomarker, the metastatic site where the cancer has spread (bone or soft tissue), and the ability of the patient to perform routine tasks (ECOG status), have been employed to assess patient risk in the metastatic prostate cancer population. For the current analysis, these conventional risk factors along with two new risk factors, circulating tumor cells (CTC) and serum testosterone, were incorporated into a proportional hazards risk model. 

A summary of the results from 631 men with complete risk factor data is provided in Table \ref{ProstateEnhanced}. All the risk factors, including CTC and serum testosterone, were prognostic for survival. To assess the adequacy of the proportional hazards assumption necessary for the interpretation of the model results, a smoothed relationship between the scaled Schoenfeld residuals and time is plotted for each risk factor \citep{Grambsch1994}. A nonconstant slope in these graphs indicates nonproportionality. The residual plots and the p-value generated from a null test statistic of a constant slope are provided in Supplemental Figure \ref{fig.phenhanced}. To achieve the proportional hazards specification in this enhanced model, the continuous risk factors, PSA, testosterone, and CTC were all transformed to a square root scale. In addition, CTC was transformed into two constructed variables based on whether the patient's CTC value exceeded 10. The diagnostic results of this model specification are consonant with the proportional hazards assumption across all risk factors. 

Although all risk factors were found prognostic, with 248 events (deaths) in the data, it is not highly informative to determine that all the log relative risk parameters are non-zero. It is often the case that after this association analysis, a second-level analysis is performed to better understand the quality of the risk model. Common families of metrics used for this evaluation include calibration, discrimination, explained variation, and likelihood based. This work will focus on the concordance probability, a measure of discrimination that gauges the extent the model-based baseline risk score separates long-term survivors from short-term survivors.  A high estimate of the concordance probability is an indication that the risk score can reliably determine patients who are at high risk and perhaps require more aggressive treatment. 

The key question we address is how to evaluate the influence of the new factors, and more generally the influence of any subset of factors, on the concordance probability in a survival model. We will use the metastatic prostate cancer example throughout the paper to illustrate the limitations of a common approach to this assessment and to propose a new solution. 


\vskip 0.50in

\section{Concordance Probability} 

For survival models, a popular discrimination measure is the concordance probability. An early definition of the concordance probability, modified to reflect the analysis horizon $\tau$, is 
\[ \mbox{Pr}[S_t(\bmX_1, \bmZ_1) <  S_t( \bmX_2, \bmZ_2)  | T_1 < T_2, T_1 < \tau]  \]
\citep{Pencina2004}, where $T$ is the survival time, $\bmX$ represents a p-dimensional vector containing the conventional risk factors, $\bmZ$ is a q-dimensional vector of new risk factors, and $S_t(\bmX, \bmZ)$ denotes the survival function $\mbox{Pr}(T > t | \bmX, \bmZ)$. Throughout the paper, upper case letters represent random variables, lower case their realizations, and bold type indicate vectors. In addition, it is assumed that at least one conventional risk factor is continuous. The concordance probability is an expression of the concordance between the survival times and the predicted survival probability beyond a given time $t$. The concordance probability ranges between 0.5 and 1.0, with the value 1 indicating perfect concordance.

The survival function $S_t(\bmX, \bmZ)$ is frequently model-based.
The most common risk models applied in clinical research are scale transformation models \citep{Cheng1995}. The scale transformation model is represented as
\[ m(T) = -\bb^T \bmX - \bg^T \bmZ + \epsilon , \]
where $m(T)$ is an unknown monotone increasing function on $[0,\tau]$, with $m(0)= -\infty$ and $m(\tau) = M < \infty$. The linear combinations  $(\bb^T\bmX, \bg^T \bmZ)$ are risk indices, and $\epsilon$ is a random error with known distribution independent of $(\bmX, \bmZ)$.

The scale transformation model may be rewritten as 
\[ g(S_T(\bmX, \bmZ)) = m(T) +\bb^T \bmX + \bg^T \bmZ ,\] 
with $\epsilon = g(S_T(\bmX, \bmZ))$ and $g(\cdot)$ is a monotone decreasing function. 
This alternative representation follows from the probability integral transform of $S$, a uniform $(0,1)$ random variable, leading to the distribution of  $\epsilon$  through the specification of $g$. Popular examples of scale transformation models include \citep{Dabrowska1988}:
\begin{enumerate}
    \item Proportional hazards : $ \ \ g(S) = \log\{- \log(S)\} $
    \item Proportional odds :  $ \ \ \ \ \ g(S) = - \log\{S/(1-S)\} $
    \item Generalized probit: $ \ \ \ \ \ g(S) = \bar{\Phi}^{-1}(S)$
\end{enumerate}
where $\bar{\Phi}$ is the standard normal survival function.

Scale transformation models contain two simplifying properties. First, there is no loss in information by considering the index survival function $S_T(\bb^T \bmX + \bg^T \bmZ)$, and second, the survival function is a monotonic decreasing function of the risk index $\bb^T \bmX + \bg^T \bmZ$. As a result, the concordance probability may be rewritten as 
\begin{equation} \kappa(\tau) = \mbox{Pr}[\bb^T\bmX_1 + \bg^T\bmZ_1 >  \bb^T\bmX_2 + \bg^T\bmZ_2 | T_1 < T_2, T_1 < \tau] ,\end{equation}
which is the consensus version of how the survival concordance probability is defined today \citep{Uno2011}. 

The most widely used estimate of the concordance probability is the concordance index \citep{Harrell1982, Harrell1996}. 
\cite{Uno2011} demonstrated that due to censoring, the concordance index (c-index) is not a consistent estimate of $\kappa(\tau)$ and incorporated an inverse probability censoring weight correction. The weighted c-index is computed as
\newline $D_n(\hat{\bb},\hat{\bg}, \hat{G}; \tau) = $
\begin{equation}  \frac{\stackunder{i, j }{\sum }  I(\hat{\bb}^T\bmx_{i} + \hat{\bg}^T\bmz_{i} > \hat{\bb}^T\bmx_j + \hat{\bg}^T\bmz_j) \ \delta_i I(y_i <y _j, y_i < \tau)   \{\hat{G}(y_i)\}^{-2}}
{\stackunder{i, j }{\sum } \delta_i I(y_i <y _j, y_i < \tau) \{\hat{G}(y_i)\}^{-2}} ,\end{equation}
\normalsize
where $\stackunder{i, j }{\sum }$ denotes a double sum taken over the $n$ observations in the sample, $Y = \min(T,C), \ \delta = I(T < C)$, and $C$ represents the underlying random censoring time with survival function $G(\cdot)$. The parameter estimates $(\hat{\bb},\hat{\bg})$ from the scale transformation model may be computed using \cite{Cheng1995}, and the Kaplan-Meier estimate of the censoring survival distribution may be used to estimate $G(y)$ under the assumption that $C \CI (T,\bmX, \bmZ)$. The weighted c-index can be generalized to include conditional independence $(C \CI T | \bmX, \bmZ)$, by replacing the Kaplan-Meier estimate with a conditional survival estimate of $G(y |\bmx, \bmz)$ \citep{Gerds2013,GerdsPEC}. When the scale transformation model is properly specified, \cite{Uno2011} provide conditions where the weighted c-index statistic is consistent and asymptotic normal.

An alternative estimate of $\kappa(\tau)$ is the concordance probability estimate \citep{Gonen2005, Zhang2018},
\begin{equation}  K_n(\hat{\bb},\hat{\bg},\hat{\theta}(\hat{\bb},\hat{\bg};\tau), \hat{\pi}; \tau) = \frac{  \stackunder{i, j }{\sum } I(\hat{\bb}^T\bmx_{i} + \hat{\bg}^T\bmz_{i} > \hat{\bb}^T\bmx_j + \hat{\bg}^T\bmz_j)  \ \hat{\theta}_{ijij}(\hat{\bb},\hat{\bg};\tau)}{ n(n-1) \hat{\pi}(\tau) } , \end{equation}
where $ \ \pi(\tau) = \mbox{Pr}[T_1 < T_2, T_1 < \tau]$ is the marginal precedence probability and 

\bigskip

\noindent $\theta_{1212}(\bb, \bg; \tau) = $ \small 
\begin{equation}  \mbox{Pr}[T_1 < T_2, T_1 < \tau | \bb^T\bmX_1=\bb^T\bmx_1, \bb^T\bmX_2=\bb^T\bmx_2,\bg^T\bmZ_1=\bg^T\bmz_1,\bg^T\bmZ_2=\bg^T\bmz_2] \end{equation} \normalsize
is the model precedence probability.  Going forward, we will simplify the notation and write $\hat{\theta}$ for $\hat{\theta}(\hat{\bb},\hat{\bg};\tau)$ in the argument of the concordance probability estimate.

The concordance probability estimate (CPE) relies on the choice of the scale transformation model used to estimate the precedence probability. For the scale transformation models introduced earlier in this section, the precedence probabilities are
\begin{itemize}

\item Proportional hazards : 
\[\theta_{1212}(\bb, \bg; \tau) = \frac{1-S_{\tau}(\bmx_1,\bmz_1)S_{\tau}(\bmx_2,\bmz_2)}{1 +\exp\left[\bb^T\bmx_{12}+\bg^T\bmz_{12}\right]}  \]
\item Proportional odds :     
\small

\noindent  $\theta_{1212}(\bb, \bg; \tau) = \exp\left[\bb^T\bmx_{12}+\bg^T\bmz_{12}\right] \ \times $
\[ \frac{\left\{\bb^T\bmx_{12}+\bg^T\bmz_{12}+\log\left(\frac{1-S_{\tau}(\bmx_2,\bmz_2)}{1-S_{\tau}(\bmx_1,\bmz_1)}\right)\right\} +\left(1-\exp\left[\bb^T\bmx_{12}+\bg^T\bmz_{12}\right]\right) \left(1-S_{\tau}(\bmx_2,\bmz_2)\right)}{ \left(1 -\exp\left[\bb^T\bmx_{12}+\bg^T\bmz_{12}\right]\right)^2} \]

\normalsize
\item Generalized probit:
\[ \theta_{1212}(\bb, \bg; \tau) = \Phi_{D,M}\left[\frac{\bb^T\bmx_{12}+\bg^T\bmz_{12}}{\sqrt{2}}, \bar{\Phi}\left(S_{\tau}(\bmx_1,\bmz_1)\right)\right] \]
\end{itemize}
where $\bmx_{12} = \bmx_1 - \bmx_2$,  $\ \bmz_{12} = \bmz_1 - \bmz_2$, and $\Phi_{D,M}$ represents a bivariate standard normal distribution function, where the first component is the difference of two independent standard normals and the second component is a marginal standard normal. The proportional hazards and proportional odds models are specific cases of the Pareto family of models. However, as studied in \cite{brentnall2018use}, the precedence probability in the general Pareto model cannot be derived analytically.

The model precedence probability estimate is a function of $(\hat{\bb}, \hat{\bg}, \hat{S}_{\tau}(\bmX, \bmZ))$. Estimates of $(\hat{\bb}, \hat{\bg})$ may be computed using Cheng et al. (1995). With $g$ assumed known, estimation of the model survival function is completed with the estimation of $m(\tau)$, which at the single point $\tau$, is attained as the solution to the estimating equation (Cheng et al. 1997)
\[ \sum_i \left\{ \frac{I(y_i \ge \tau)}{\hat{G}(\tau)} - g^{-1}[m(\tau) + \hat{\bb}^T\bmx_i +\hat{\bg}^T\bmz_i] \right\} = 0  .\]

\bigskip
The CPE in (3) is a consistent, asymptotic normal estimate of $\kappa(\tau)$ when the scale transformation model is properly specified \citep{Devlin2021}.  Its consistency is apparent by applying Bayes Theorem in (1) to obtain

\begin{equation} \kappa(\tau) = [\pi(\tau)]^{-1} \stackunder{\beta^T x_{12} + \gamma^T z_{12} >  0}{\int \int \int \int} \theta_{1212}(\bb, \bg; \tau) dF(\bb^T\bmx_1, \bb^T\bmx_2,  \bg^T\bmz_1, \bg^T\bmz_2) . \end{equation}

The CPE does not require an inverse probability censoring weight, making it less sensitive to censoring in the right tail than the weighted c-index, and is more efficient than the weighted c-index under the correct model specification. Thus, the CPE is advantageous to apply under proper model specification. However, if the enhanced model is incorrectly specified, the weighted c-index combined with the partial rank estimate (Section 3.3), is more robust to model misspecification and has greater efficiency than the CPE.  These assertions are supported by the simulations in Section 4.

\vskip 0.50in

\section{Impact of New Factors on the Concordance Probability} 

{\it 3.1 Nested Models Approach}

\noindent To determine the impact of a new set of factors on the concordance probability, the current practice is to use the estimates from nested scale transformation models 
\begin{equation} m(T) = -\bb^T \bmX - \bg^T \bmZ +\epsilon \end{equation}
\begin{equation} m^*(T) = -\bb^{*T} \bmX + \epsilon^* \end{equation}
and calculate the difference in the estimated concordance probabilities computed from the enhanced model (6) and the reduced model (7). The expectation is that the addition of new factors to the model will improve the understanding of patient risk. 

For the metastatic prostate cancer data, the results from the enhanced model were presented in Section 1. The results for the reduced proportional hazards model, based on the conventional risk factors, PSA, visceral disease, and ECOG status, are presented in Table \ref{ProstateReduced}. The conventional risk factors remain prognostic; however, the proportional hazards assumption does not hold for visceral disease, as illustrated in Supplemental Figure \ref{fig.phreduced}. This scenario, where the proportional hazards assumption is satisfied in the investigator-constructed enhanced model, but violated in the reduced model is unsurprising, since proportional hazards models are, in general, non-nested \citep{Hougaard1986, Fine2002}. Thus, a limitation of comparing estimates of the concordance probabilities from nested scale transformation models is the structural misspecification of the reduced model due to the non-nesting property of proportional hazards models. This misspecification will affect the estimated concordance probability from the reduced model using either the weighted c-index or CPE and, ultimately, the interpretation of the impact of circulating tumor cells and serum testosterone on this discrimination measure.  

The non-nesting argument for the proportional hazards assumption may be generalized to the family of scale transformation models.  If the heterogeneity due to $\bmZ$ in the enhanced scale transformation model (6)
is ignored by incorporating $\bg^T \bmZ$ into the error term, the reduced model is written as (7),
where $\epsilon^*$ is a location-scale transformation of $\epsilon - \bg^T \bmZ$. In general, the distribution of $\epsilon^*$ is not the same as $\epsilon$, and hence the scale transformation family models are nonnested. The generalized probit model is an exception if the distribution of $(\bmX, \bmZ)$ is multivariate normal.

A measure for the impact of the new factors to the concordance probability, based on the nested models in (6) and (7),  is 
\small 
\[  \mbox{Pr}[\bb^{T}\bmX_1 + \bg^T \bmZ_1 >  \bb^{T}\bmX_2 + \bg^T \bmZ_2  | T_1 < T_2, T_1 < \tau] - \mbox{Pr}[\bb^{* T}\bmX_1  >  \bb^{* T}\bmX_2  | T_1 < T_2, T_1 < \tau] .\]
\normalsize
This metric is a function of the relationship between the new risk factors $\bmZ$ and survival time $T$ in the presence of the conventional risk factors $\bmX$ and the degree of misspecification of the reduced model.  It is clearly undesirable for the influence of the new factors on the concordance probability to be a function of the level of misspecification. Thus, we will not proceed further with this nested model approach.

In the next two sections, alternative approaches that do not rely on the misspecified reduced model but instead use a projection approach, are provided to compute the impact of the new factors on the concordance probability. In Section 3.2, the methodology is applied when the error distribution for $\epsilon$ from a scale transformation model is known, and Section 3.3 considers the case that the error distribution from this model is unknown. Estimation for the concordance probability and inference for the impact parameter are carried out separately depending on the error distribution information.  

\vskip 0.20in

\noindent {\it 3.2 Error Distribution Known}

\noindent  To develop a concordance probability measure that ignores the heterogeneity due to the new factor risk index $\bg^T \bmZ$, the enhanced model precedence probability (4) is projected onto the two-dimensional space spanned by the random variables $(\bb^T \bmX_1, \bb^T \bmX_2)$,
\begin{equation} \theta_{12}^{[P]}(\bb; \tau) = E\left[ \theta_{1212}(\bb, \bg;\tau) \big| \bb^T\bmX_1, \bb^T\bmX_2 \right] , \end{equation}
which as a direct result of iterated expectation
\[    \theta_{12}^{[P]}(\bb; \tau) = \mbox{Pr}[T_1 < T_2, T_1 < \tau | \bb^T \bmX_1, \bb^T \bmX_2] . \]
Through this projected precedence probability, the concordance probability, ignoring the new factors in the enhanced model, is defined in an analogous way to (5) as
\begin{equation} \kappa^{[P]}(\tau) = [\pi(\tau)]^{-1} \stackunder{\beta^T x_{12} >  0}{\int \int}  \theta_{12}^{[P]}(\bb; \tau) dF(\bb^T\bmx_1, \bb^T\bmx_2) .\end{equation}
which equals 
\begin{equation}\mbox{Pr}[\bb^T \bmX_1 > \bb^T \bmX_2 | T_1 < T_2, T_1 < \tau]. \end{equation}

The concordance probability from the enhanced model $\kappa(\tau)$, and the concordance probability for the conventional factor risk index alone $\kappa^{[P]}(\tau)$, may each be interpreted as projections. $\kappa(\tau)$ is proportional to the projection of the random variable $I(T_1 < T_2, T_1 < \tau)$ onto the half-space $\bb^T\bmX_{12} + \bg^T\bmZ_{12}  > 0$ and $\kappa^{[P]}(\tau)$ is proportional to the projection of this random variable onto the half-space $\bb^T\bmX_{12}  >0$. The difference in the concordance probabilities leads to a measure of the impact of the new factors in the enhanced model
\begin{equation}  \xi(\tau) = \kappa(\tau) - \kappa^{[P]}(\tau) .\end{equation}

The efficiency of the CPE relative to the weighted c-index when the error distribution in (6) is known, suggests that 
the estimated impact parameter be computed as 
\[ \hat{\xi}(\tau) = K_n(\hat{\bb}, \hat{\bg}, \hat{\theta},\hat{\pi} ; \tau) - K_n^{[P]}(\hat{\bb}, \hat{\theta}^{[P]},\hat{\pi} ; \tau) ,\]
 where $K_n(\hat{\bb}, \hat{\bg},\hat{\theta},\hat{\pi} ; \tau)$ is expressed in (3) and  
\[ K_n^{[P]}(\hat{\bb}, \hat{\theta}^{[P]},\hat{\pi}; \tau) = \frac{\stackunder{i, j}{\sum} I[\hat{\bb}^T\bmx_{i}  > \hat{\bb}^T\bmx_{j}]  \hat{\theta}_{ij}^{[P]}(\hat{\bb}; \tau)}{n(n-1) \ \hat{\pi}(\tau)} .\]
The projected precedence probability $\theta_{12}^{[P]}(\bb;\tau)$, a conditional expectation (8), is estimated using kernel smoothing
\[ \hat{\theta}_{12}^{[P]}(\hat{\bb}; \tau) = \frac{ \stackunder{k,l}{\sum} \hat{\theta}_{12kl}(\hat{\bb}, \hat{\bg};\tau) \phi_h(\hat{\bb}^T\bmx_1,\hat{\bb}^T\bmx_k) \phi_h(\hat{\bb}^T\bmx_2,\hat{\bb}^T\bmx_l)}
 {\stackunder{k,l}{\sum} \phi_h(\hat{\bb}^T\bmx_1,\hat{\bb}^T\bmx_k) \phi_h(\hat{\bb}^T\bmx_2,\hat{\bb}^T\bmx_l)}  ,\]
where $\phi_h(u,v)$ is a kernel function with bandwidth $h$. The justification for the projected precedence probability estimate is provided in Theorem 1.
 
\vskip 0.30in

\noindent {\bf Theorem 1:}
\newline Let $\phi_h(u,v)$ represent a kernel function, symmetric about zero, with bandwidth $h$. Assume as $n \rightarrow \infty$, $h \rightarrow 0$ and $nh^2 \rightarrow \infty$. Then
 \[ \hat{\theta}_{12}^{[P]}(\hat{\bb}; \tau) \ \stackrel{p}{\rightarrow} \  \mbox{Pr}(T_1 < T_2, T_1 < \tau| \bb^T\bmX_1, \bb^T\bmX_2)  \] 
Theorem 1  is derived in the appendix and follows the conventional asymptotic result using a Nadaraya-Watson estimator \citep{Simonoff1996}. 

The asymptotic distribution of $\hat{\xi}(\tau)$ is developed to obtain its asymptotic variance and to produce a 95\% confidence interval for $\xi(\tau)$. These inferential measures will be applied in Section 5 to complete the prostate cancer data analysis. The asymptotic distribution for the CPE impact estimate is provided in Theorem 2 and is derived in the appendix.

\vskip 0.35in

\noindent {\bf Theorem 2:}
\newline Assume the scale transformation model based on the covariates $(\bmx, \bmz)$ is properly specified, and the new factors are associated with survival time
$(\bg \ne 0)$. Then 

\begin{equation} n^{1/2} [ K_n(\hat{\bb},\hat{\bg}, \hat{\theta},\hat{\pi}; \tau) - K_n^{[P]}(\hat{\bb}, \hat{\theta}^{[P]},\hat{\pi} ; \tau)  - \xi(\tau)]  \ \stackrel{D}{\rightarrow} \ N(0, V_K) . \end{equation}

\vskip 0.35in

The result is derived by decomposing (12) into components and demonstrating the asymptotic normality of each component. An analytic estimate of the asymptotic variance $V_K$ via the sum of these components is complex, and a bootstrap estimate of the asymptotic variance and confidence interval will be applied \citep{Kosorok2004}.

 \vskip 0.35in

\noindent {\it 3.3 Error Distribution Unknown}

In the previous section, the impact of the new factors on the concordance probability was developed under the assumption that the error distribution was known.  The working premise is that the analyst has carefully evaluated the data and chosen from the family of enhanced scale transformation models in Section 2.
However, there will be times when even a close approximation will not be suitable. 

For these data, estimates of $(\bb, \bg)$ may be derived, without specifying an error distribution for $\epsilon$, by using the partial rank methodology (Khan and Tamer, 2007). The partial rank estimates (PRE) of  $(\hat{\bb},\hat{\bg})$ are obtained from the objective function
\[ J_n(\bn, \bg) = \argmax_{(\bn,\bg)} \ \ [n(n-1)]^{-1} \sum_i \sum_j \delta_j I[y_i > y_j] I[\bb^T \bmx_i + \bg^T \bmz_i < \bb^T \bmx_j + \bg^T \bmz_j] ,\]
where to resolve an identifiability issue in this maximization, 
the first component of $\bb$ is set to one, and $\hat{\bb} = (1, \hat{\bn}^T)^T$.  \cite{Khan2007} prove that, under regularity conditions, which includes that there is at least one continuous covariate with a non-zero coefficient,  $(\hat{\bn}, \hat{\bg})$ are asymptotically normal and $(\hat{\bb}, \hat{\bg})$ are consistent estimates of the rescaled parameters $(\bb, \bg)$ in (6). However, the discontinuity in the objective function can result in instability in this estimation process. To stabilize the algorithm, a smooth version of the objective function is employed
\[ \tilde{J}_n(\bn, \bg) = \argmax_{(\bn,\bg)} \ \ [n(n-1)]^{-1} \sum_i \sum_j \delta_j I[y_i > y_j] \Phi \left( \frac{\bb^T(\bmx_j - \bmx_i) + \bg^T(\bmz_j - \bmz_i)}{g} \right) , \]
where $\Phi(\cdot)$ represents a local normal distribution function and $g$ is its bandwidth which converges to 0 as $n$ increases \citep{song2007semiparametric}.

With the coefficient estimates $(\hat{\bb}, \hat{\bg})$, the impact of the new factors on the weighted c-index may be estimated by
\[  D_n(\hat{\bb}, \hat{\bg}, \hat{G} ; \tau) - D_n^{[P]}(\hat{\bb}, \hat{G} ; \tau) \]
 where $D_n(\hat{\bb}, \hat{\bg}, \hat{G} ; \tau)$ is expressed in (2), and from (10), 
 \[ D_n^{[P]}(\hat{\bb}, \hat{G} ; \tau) = \frac{\stackunder{i, j }{\sum } I[\hat{\bb}^T\bmx_i > \hat{\bb}^T\bmx_j] \delta_i I[y_i <y _j, y_i < \tau] \{\hat{G}(y_i)\}^{-2}} 
{\stackunder{i, j }{\sum } \delta_i I[y_i <y _j, y_i < \tau] \{\hat{G}(y_i)\}^{-2}},  \]
is a consistent estimate of the projected concordance probability $\kappa^{[P]}(\tau)$.  In contrast to Section 3.2, the weighted c-index is employed because the model precedence probability cannot be computed without specifying the error distribution. The use of the weighted c-index simplifies the calculation of the impact estimate, but at the cost of including inverse probability censoring weights. The asymptotic distribution for the estimated impact, using the weighted c-index, is provided in Theorem 3.   

\vskip 0.35in

\noindent {\bf Theorem 3:}
\begin{equation} n^{1/2} [ D_n(\hat{\bb},\hat{\bg}, \hat{G}; \tau) - D_n^{[P]}(\hat{\bb}, \hat{G}; \tau)  - \xi(\tau)]  \ \stackrel{D}{\rightarrow} \ N(0, V_D) . \end{equation}

\vskip 0.35in

\noindent This result follows directly by noting that 

\bigskip

\noindent $D_n(\hat{\bb}, \hat{\bg}, \hat{G} ; \tau) - D_n^{[P]}(\hat{\bb}, \hat{G} ; \tau) = $
\[ \frac{\stackunder{i, j }{\sum } \left[ I(\hat{\bb}^T\bmx_{i} + \hat{\bg}^T\bmz_{i} > \hat{\bb}^T\bmx_j + \hat{\bg}^T\bmz_j) - I(\hat{\bb}^T\bmx_{i} > \hat{\bb}^T\bmx_j) \right]  \delta_i I(y_i <y _j, y_i < \tau)   \{\hat{G}(y_i)\}^{-2}}
{\stackunder{i, j }{\sum } \delta_i I(y_i <y _j, y_i < \tau) \{\hat{G}(y_i)\}^{-2}} \]
and applying the asymptotic normality derivation in \cite{Uno2011} to the statistic above.

\vskip 0.50in

\section{Simulation Studies} 

Simulations were performed to estimate the adequacy of the projection estimate of the concordance probability and the resultant impact estimate. The three approaches detailed in this work were considered.
\begin{enumerate}
    \item  The enhanced model coefficients were estimated with the partial likelihood, and the concordance probabilities were computed with the CPE. [PL/CPE]
    \item The enhanced model coefficients were estimated with the partial likelihood, and the concordance probabilities were estimated with the weighted c-index. [PL/wCI]
    \item The enhanced model coefficients were estimated using the partial rank method, and the concordance probabilities were computed with the weighted c-index. [PR/wCI] 
    \end{enumerate}
The abbreviations in square brackets will be used freely throughout this section to denote the three approaches.

The first set of simulations evaluated the estimates under proportional hazards, while the second evaluated the estimates under non-proportional hazards. Across all simulations, two conventional factors, $X_1$ and $X_2$, and two new factors under evaluation, $Z_1$ and $Z_2$, were generated as independent normal random variables with mean 0 and variance 1. All four factors generated the underlying survival times for the enhanced model, and the censoring times were generated independently of all factors. Impact parameters of 0.025, 0.05, and 0.10 for the concordance probability were evaluated. Across all scenarios, the concordance probability of the enhanced model was 0.70.

For the proportional hazards simulations, survival times were generated using the regression model $t_i =\exp\{ -(\beta_1x_{i1}+\beta_2 x_{i2}+\gamma_1 z_{i1}+\gamma_2 z_{i2})\} \times \epsilon_i$, where $\epsilon_i$ were independent exponential random variables with a scale parameter equal to 1. To achieve an impact $\xi(\tau)$ of 0.025 on the concordance probability from the enhanced model, the following parameters were selected: $\beta_1=0.718$, $\beta_2=0.15$, $\gamma_1=0.346$, $\gamma_2=0.15$, and using $\tau=1.18$. Censoring times were generated from a Uniform($0$, $b$), where $b$ was selected to be  1.58 and 4.75 to achieve an average censoring rate of 50\% and 25\%, respectively. The parameters were also estimated when the censoring rate was 0\%, and all failure times were observed. To achieve an impact $\xi(\tau)$ of 0.05, the following parameters were selected: $\beta_1=0.624$, $\beta_2=0.15$, $\gamma_1=0.505$, $\gamma_2=0.15$. Lastly, for an impact of 0.10, the parameters were $\beta_1=0.408$, $\beta_2=0.15$, $\gamma_1=0.684$, and $\gamma_2=0.15$. The value of $\tau$ and the two censoring distributions remained the same.

Under non-proportional hazards, survival times were generated from a conditional proportional hazards model with a gamma frailty, $t_i = \exp\{-(\beta_1x_{i1}+\beta_2 x_{i2}+\gamma_1 z_{i1}+\gamma_2 z_{i2} + \log w_i) \} \times \epsilon_i$, where $w_i$ are independent gamma random variables with shape and scale parameters equal to 0.25, and the $\epsilon_i$ were independent exponential random variables with a scale parameter of 1. To achieve an impact of 0.025 in the concordance probability, the following parameters were selected: $\beta_1= 1.741$, $\beta_2=0.15$, $\gamma_1=0.887$, and $\gamma_2=0.15$, along with $\tau=6$. Censoring times were again generated from Uniform($0$, $b$), where the values of 13 and 310 for $b$ achieved an average censoring rate of 50\% and 25\%, respectively, in addition to the scenario when the rate was 0\%. For an impact of 0.05, the parameters were $\beta_1=1.567$, $\beta_2=0.15$, $\gamma_1=1.301$, and $\gamma_2=0.15$.  For an impact of 0.10, the parameters were $\beta_1=1.061$, $\beta_2=0.15$, $\gamma_1=1.754$, and $\gamma_2=0.15$. The values of $\tau$ and $b$ remained the same. 

For each iteration of the data-generating process, the impact of the new risk factors $\xi(\tau)$ on the concordance probability from the enhanced model, was estimated using the CPE projection framework via the estimates $K_n(\hat{\bb},\hat{\bg}, \hat{\theta},\hat{\pi}; \tau)$ and $K_n^{[P]}(\hat{\bb}, \hat{\theta}^{[P]},\hat{\pi}; \tau)$ and with the weighted c-index using $D_n(\hat{\bb}, \hat{\bg}, \hat{G}; \tau)$ and $D_n^{[P]}(\hat{\bb}, \hat{G}; \tau)$ using the proportional hazards (PH) model estimates or the partial rank estimates (PR).  Across 2,000 iterations, these approaches were compared based on the average bias and the relative efficiency using the root mean square error (rMSE) of the CPE with respect to the weighted c-index using either PH or PR. In addition, for each iteration of the data-generating process, the respective standard errors for each concordance measure were estimated using 50 bootstrap samples. Using these standard error estimates, the ratio of the estimated standard error to the simulation standard error was calculated along the 95\% coverage probability. The sample size for all simulations was 300. The impact and projected concordance probability parameters were determined by simulation from the average of 2,000 iterations, with each iteration containing 2,000 uncensored observations.

Results for the proportional hazards simulations are presented in Tables \ref{PHsimresultsDelta0025}, \ref{PHsimresultsDelta005}, and \ref{PHsimresultsDelta010} when the impact parameter in the concordance probability was 0.025, 0.05, and 0.10, respectively, based on the two new risk factors. For the three approaches, the average biases for the projection and impact estimates were minimal across scenarios. Using the bootstrap-based standard error estimates, all methods provided coverage near the 95\% nominal level for both parameters, and the ratios of the estimated standard error to the simulation standard error were aligned close to 1. Under correct specification of the underlying model, the PL/CPE had relative efficacy gains of 10-31\% over the PL/wCI and the PR/wCI, where the gains depended on the censoring rate and whether the projection concordance probability or impact parameter was being estimated. 

Tables \ref{NPHsimresultsDelta0025}, \ref{NPHsimresultsDelta005}, and \ref{NPHsimresultsDelta010} provide the simulation results under the non-proportional hazards scenario when the impact on concordance was 0.025, 0.05, and 0.10, respectively. Due to the underlying model misspecification, the PL/CPE approach produced biased estimates for both the projection concordance and impact parameters, along with poor coverage. Importantly, the test for the proportional hazards assumption \citep{Grambsch1994} was rejected 100\% when there was no censoring and at least 93\% of the time across all scenarios (Supplemental Table, \ref{simPHtest}).  This reiterates the need to interrogate any modeling assumptions before selecting the methodological approach to estimate the projection and impact parameters. While the average bias was small for the PL/wCI estimates, the simulation standard errors were misaligned with the estimated bootstrap standard errors in some scenarios. These situations led to inefficiencies and weakened coverage relative to the PR/wCI estimates.

These findings imply that in cases where the data closely align with the proportional hazards assumption, the partial likelihood/CPE projection approach outperforms in estimating the influence of new risk factors on the concordance probability. This superiority is attributed to the generation of smaller standard error estimates and, consequently, the formation of more tightly bound 95\% confidence intervals.  However, when the data are not well approximated by  proportional hazards, or more generally, the models in Section 2, the partial rank estimate/weighted c-index is the preferred approach. 

\vskip 0.50in

\section{Prostate Cancer Example - Conclusion} 

Six hundred and thirty one men with complete risk factor data were treated with an androgen receptor targeted therapy as part of a randomized clinical trial in metastatic prostate cancer. The objective was to use the survival and risk factor data to evaluate the impact of the two new biomarkers on the concordance probability. The rationale for the inclusion of the new biomarkers, circulating tumor cells and serum testosterone, are distinct.  Circulating tumor cells provide a blood-based quantification of tumor burden with high levels a direct measure of advanced disease.  In contrast, low values of serum testosterone in the presence of metastatic disease is an indirect indication that the prostate cancer is aggressive.
 
The maximum event time in this cohort was 32.7 months, which led to the choice $\tau = 30$ months for the follow-up time used in the concordance probability analysis. The CPE for the enhanced model was 0.704 and the $95\%$ confidence interval for the concordance probability parameter (1) was $(0.679-0.729)$. Applying the projection method in Section 3.2, the CPE estimated impact parameter, computed from $K_n(\hat{\bb},\hat{\bg},\hat{\theta}, \hat{\pi}; \tau)$ and  $K_n^{[P]}(\hat{\bb}, \hat{\theta}^{[P]},\hat{\pi} ; \tau)$, was 0.088, and the $95\%$ confidence interval for the impact parameter (11) was $(0.060-0.115)$. 
The weighted c-index for the enhanced model using the partial rank estimate was $0.735$, with its attendant $95\%$ confidence interval $(0.695-0.775)$.  
The weighted c-index produced a comparable impact estimate, computed from $D_n(\hat{\bb},\hat{\bg}, \hat{G}; \tau)$ and $D_n^{[P]}(\hat{\bb}, \hat{G} ; \tau)$, equal to 0.095, and the $95\%$ confidence interval for the impact parameter was $(<0.01-0.190)$. As anticipated, the confidence interval using the partial rank estimate was wider than the CPE-based interval, which results from the improved efficiency of the CPE when the data are well approximated by proportional hazards. 

Historically, PSA has been one of the foremost prognostic serum biomarkers employed in prostate cancer research. When subtracting the standard CPEs from two nested incorrectly specified proportional hazards models, namely (i) ECOG status and visceral disease presence, and (ii) PSA, ECOG status, and visceral disease presence, the impact of removing PSA from the conventional risk model resulted in a CPE reduction of  $-0.019$  (95\% CI: -0.046, 0.009).  This suggests either no impact or a marginal decrease in PSA's prognostic impact  on the CPE among the conventional risk factors, and contradicts the conventional wisdom in the field. The proportionality of the enhanced model including CTC and serum testosterone, however, may be leveraged to employ the projection method to gauge the impact of PSA on the CPE among conventional factors. For this purpose, the impact of CTC and testosterone on the enhanced model was subtracted from the impact of PSA, CTC, and testosterone on the enhanced model. The difference was 0.031 (95\% CI: 0.008, 0.054), supporting the positive influence of PSA on the CPE within the conventional risk factors.  

These results provide unequivocal support for incorporating PSA, circulating tumor cells and testosterone into the enhanced risk model. However, the estimates of the concordance probability from the enhanced model containing all the risk factors (0.704 and 0.735) indicate that further research is needed to generate new risk factors in this population before the model may be applied to assist in clinical decisions.

\vskip 0.50in

\section{Discussion} 

\noindent In biology and medicine, the discovery of novel biological factors is often followed by an assessment of their association with one or a set of clinical outcomes, a key step as a discovery goes from the bench to the bedside.  If an association is established,  subsequent empirical research is needed to evaluate the impact of these novel factors relative to known risk factors in the patient population.  This step, which often requires a model to link the risk factors to the clinical outcome, is useful in determining where the discovery lies on the spectrum between redundant and orthogonal risk factor information.

The value of a risk model may be partly summarized using measures such as calibration, discrimination, likelihood-based, and explained variation. In this work, we focused on the concordance probability, a discrimination metric that synthesizes well when applied to scale transformation models. Naively, one approach to evaluate the influence of the new risk factors is to compute the concordance probability in a model containing both conventional risk factors and new risk factors and compare it to the concordance probability using the conventional risk factors alone. The issue with this frequently utilized approach stems from the fact that scale transformation models, in general, do not exhibit nesting. This lack of nesting may subsequently affect estimation of the influence of new factors on the concordance probability.

To ascertain the impact of the new risk factors on the concordance probability, a new approach using projection methodology was developed with the concordance probability estimate (CPE) and the weighted c-index. This approach does not apply a reduced model, enabling the analyst to focus on the discriminative aspects of the newly introduced risk factors within the enhanced survival model. The choice of concordance estimates was dependent on whether the analyst was confident that the data fit one of the specified models in Section 2. The most commonly applied model in this family is the proportional hazards model. One benefit of identifying the error distribution (Section 3.2) is greater efficiency of the impact statistic through the use of the CPE. An additional benefit is the interpretability of the parameters $(\bb, \bg)$ in the scale transformation model. As noted in \cite{Dabrowska1988}, the interpretation of the parameters for the models considered in Section 2 are independent of the transformation $m(\cdot)$. 

The CPE approach assumes that the analyst has a good understanding of their data and can identify an enhanced scale transformation model to approximate this data. In practice, this may require covariate stratification, truncation of the follow up time, or the transformation of covariates. In the prostate data, the effect of circulating tumor cells (CTCs) was bifurcated based on whether the baseline value was less than or greater than ten CTCs to approximate a proportional hazards specification. There will be times, however, when one cannot approximate a properly specified enhanced scale transformation model. This includes the special case where it is the reduced model that can be properly specified. In these cases, it is recommended that the analyst use the partial rank methodology to estimate the coefficients from the enhanced model and apply the weighted c-index along with its projection to evaluate the impact of the new covariates. One alternative to the partial rank estimate in this scenario,  is to find a superset of factors that are properly specified by a scale transformation model, and evaluate the two subsets of risk factors using the projection method. This was the method we used to determine the impact of PSA among the conventional risk factors. A second method suggested by a reviewer, is to consider nonparametric maximum likelihood estimation (Zeng and Lin, 2007), which requires joint estimation of the infinite dimensional parameter $m(t)$ and the finite dimensional coefficient parameters $(\bb, \bg)$. 

For the prostate cancer data in this paper, the concordance probability estimate, measured on a scale between 0.5 and 1.0, was only 0.70 in the enhanced model, an indication that further research is needed to find new factors that discriminate between longer-term survivors and short-term survivors. The analysis also demonstrated that the new biomarkers CTC and serum testosterone were highly influential in discriminating patient risk. The benefit of CTC alone can be further magnified using the proposed projection methodology. Ignoring the heterogeneity due to CTC in the enhanced risk model, produced a much smaller CPE equal to 0.62. Alternatively, ignoring all factors but CTC in the risk model, produced a concordance probability estimate equal to 0.67, surprisingly close to the enhanced model CPE.  Thus, although the CTC assay costs more than the combined costs of the PSA and serum testosterone assay, there is a benefit to its inclusion in the risk model. Further research evaluating  a cost-benefit analysis for CTC and overall survival would provide additional useful information. 

The impact measure was derived from a scale transformation survival model. The benefit of the scale transformation model is the reduction in dimensionality in the risk assessment through a single risk index. The application of a single index model is justified as an approximation to a more complex risk function \citep{Hall1989}.  Alternative approaches to understanding individual risk, which explicitly encompasses greater risk complexity and dimensionality, include random forest, neural network, and nonparametric regression models.  Evaluation of the impact of new factors in this high dimensional high complexity case will be the subject of future research.

\section*{Code Availability}
The code to implement the projection-based impact parameter is available using the R package `SurvEval,' which is on CRAN \citep{DHSurvEval}.


\clearpage
\newpage
\bibliography{References.bib}

\clearpage
\newpage

\noindent \underline{Tables and Figures}

\begin{table}[ht] 
\centering
 \caption{Proportional hazards estimates for the enhanced model with 631 metastatic prostate cancer patients for the endpoint of overall survival. A total of 248 died during the follow-up period.}  
\begin{tabular}{rrrrrr} \label{ProstateEnhanced} 
 & log(RR)  & se[log(RR)] & $p$-value \\ 
  \hline
PSA & 0.018  & 0.006 & 0.002 \\ 
Visceral Disease & 0.387 & 0.156 &  0.013 \\ 
  ECOG Status & -0.408  & 0.134 &  0.002 \\ 
  Serum Testosterone & -1.625  & 0.453 & $<$0.001 \\ 
  CTC$_{\text{low}}$ & 0.258  & 0.076 & 0.001 \\ 
    CTC$_{\text{high}}$  & 0.106 & 0.009 & $<$0.001 \\ 
   \hline
   \multicolumn{5}{l}{PSA, testosterone, and CTC were estimated under a square root} \\
     \multicolumn{5}{l}{transformation. CTC$_{\text{low}}$ are the transformed values $< \sqrt{10}$, 0 otherwise, } \\
      \multicolumn{5}{l}{ while CTC$_{\text{high}}$ are values $\geq \sqrt{10}$, 0 otherwise. RR is the relative risk. } 
   \end{tabular}
\end{table}

\begin{table}[ht]
\centering
 \caption{Proportional hazards estimates for the reduced working model.}  
\begin{tabular}{rrrrrr} 
  \hline
 & log(RR)  & se[log(RR)] & $p$-value \\
  \hline
PSA & 0.030 &   0.004 & $<$0.001 \\ 
Visceral Disease & 0.378 & 0.154 & 0.014 \\ 
  ECOG Status  & -0.484 &  0.132   &$<$0.001 \\ 
   \hline
      \multicolumn{5}{l}{PSA was estimated under a square root transformation.} \\
\end{tabular} \label{ProstateReduced}
\end{table}

\clearpage

 \begin{table}[ht]
 \caption{Proportional hazards simulation results for the impact parameter equal to 0.025.
  The average bias and relative efficiency of the projection-based CPE with respect to the weighted c-index were averaged over 2,000 simulation iterations. Standard errors were calculated based on 50 bootstrap samples within each iteration and compared to the simulation standard error. Coverage was based on 95\% confidence intervals using the bootstrap standard error estimate. }
\centering
\tiny
\begin{tabular}{lccccccccccccc} \label{PHsimresultsDelta0025}
\vspace{-0.25cm}
   &&&&&& & Relative Efficiency  & Estimated to  &\\
& N & Enhanced &  Projection & Estimate & Censoring &Bias & CPE w.r.t. wCI & Simulation SE & Coverage \\
 \hline
 \hline
  Projection &&&&&&&&& \\
& 300 &0.70 & 0.675 & PL/CPE & 0\% &  0.001 & 1.000 & 0.991 & 0.949 \\  
& 300 &0.70 & 0.675 & PL/wCI &0\%  &  <0.001 & 1.225 & 1.010 & 0.951 \\
& 300 &0.70 & 0.675 & PR/wCI &0\%  &  <0.001 & 1.225 & 1.010 & 0.951 \\ 
\hline
& 300 &0.70 & 0.675 & PL/CPE  &25\% &   0.001 & 1.000 & 0.999 & 0.953 \\ 
& 300 &0.70 & 0.675 & PL/wCI &25\%  &  0.000 & 1.164 & 1.013 & 0.954 \\ 
& 300 &0.70 & 0.675 & PR/wCI &25\%  &   0.001 & 1.165 & 1.013 & 0.952 \\ 
\hline
& 300 &0.70 & 0.675 & PL/CPE  &50\%  &  0.001 & 1.000 & 0.972 & 0.939 \\ 
& 300 &0.70 & 0.675 & PL/wCI &50\%  &   0.001 & 1.094 & 1.003 & 0.953 \\
& 300 &0.70 & 0.675 & PR/wCI &50\%  &  0.001 & 1.096 & 1.007 & 0.951 \\ 
  \hline
  \hline
  Impact &&&&&&&&& \\
& 300 &0.70 & 0.675 & PL/CPE   &0\%  &  0.001 & 1.000 & 1.006 & 0.948 \\ 
& 300 &0.70 & 0.675 & PL/wCI &0\%  &   0.001 & 1.324 & 1.028 & 0.957 \\ 
& 300 &0.70 & 0.675 & PR/wCI &0\%  &   0.002 & 1.344 & 1.034 & 0.953 \\ 
\hline
&   300 &0.70 & 0.675 & PL/CPE   &25\% & 0.002 & 1.000 & 1.014 & 0.949 \\ 
& 300 &0.70 & 0.675 & PL/wCI &25\%  &  0.002 & 1.212 & 1.040 & 0.958 \\ 
& 300 &0.70 & 0.675 & PR/wCI &25\%  &  0.002 & 1.227 & 1.045 & 0.954 \\ 
\hline
& 300 &0.70 & 0.675 & PL/CPE   &50\%  & 0.003 & 1.000 & 1.019 & 0.946 \\ 
& 300 &0.70 & 0.675 & PL/wCI &50\% & 0.003 & 1.108 & 1.072 & 0.959 \\
& 300 &0.70 & 0.675 & PR/wCI &50\% &   0.003 & 1.116 & 1.081 & 0.961 \\
  \hline
  \hline
\end{tabular}
\end{table}

 \begin{table}[ht]
 \caption{Proportional hazards simulation results for the impact parameter equal to 0.05.
  The average bias and relative efficiency of the projection-based CPE with respect to the weighted c-index were averaged over 2,000 simulation iterations. Standard errors were calculated based on 50 bootstrap samples within each iteration and compared to the simulation standard error. Coverage was based on 95\% confidence intervals using the bootstrap standard error estimate. }
\centering
\tiny
\begin{tabular}{lccccccccccccc} \label{PHsimresultsDelta005}
\vspace{-0.25cm}
   &&&&&& & Relative Efficiency  & Estimated to  &\\
& N & Enhanced &  Projection & Estimate & Censoring &Bias & CPE w.r.t. wCI & Simulation SE & Coverage \\
 \hline
 \hline
  Projection &&&&&&&&& \\
& 300 &0.70 & 0.65 & PL/CPE & 0\% & 0.002 & 1.000 & 1.040 & 0.959 \\ 
& 300 &0.70 & 0.65 & PL/wCI &0\%  & 0.001 & 1.238 & 1.018 & 0.951 \\ 
& 300 &0.70 & 0.65 & PR/wCI &0\%  & 0.001 & 1.238 & 1.018 & 0.954 \\  
\hline
& 300 &0.70 & 0.65 & PL/CPE  &25\% & 0.002 & 1.000 & 1.003 & 0.950 \\ 
& 300 &0.70 & 0.65 & PL/wCI &25\%  & 0.001 & 1.176 & 1.012 & 0.950 \\ 
& 300 &0.70 & 0.65 & PR/wCI &25\%  & 0.001 & 1.177 & 1.014 & 0.949 \\ 
\hline
& 300 &0.70 & 0.65 & PL/CPE  &50\%  & 0.002 & 1.000 & 0.980 & 0.941 \\ 
& 300 &0.70 & 0.65 & PL/wCI &50\%  & 0.002 & 1.101 & 1.008 & 0.949 \\ 
& 300 &0.70 & 0.65 & PR/wCI &50\%  &  0.002 & 1.104 & 1.013 & 0.952 \\
  \hline
  \hline
  Impact &&&&&&&&& \\
&    300 &0.70 & 0.65 & PL/CPE   &0\%  &0.001 & 1.000 & 1.020 & 0.956 \\ 
& 300 &0.70 & 0.65 & PL/wCI &0\%  &  0.001 & 1.304 & 1.019 & 0.956 \\ 
& 300 &0.70 & 0.65 & PR/wCI &0\%  &  0.002 & 1.314 & 1.021 & 0.957 \\ 
\hline
&   300 &0.70 & 0.65 & PL/CPE   &25\%  &0.001 & 1.000 & 1.012 & 0.949 \\ 
& 300 &0.70 & 0.65 & PL/wCI &25\%  & 0.001 & 1.230 & 1.015 & 0.954 \\ 
& 300 &0.70 & 0.65 & PR/wCI &25\%  & 0.002 & 1.240 & 1.016 & 0.955 \\ 
\hline
& 300 &0.70 & 0.65 & PL/CPE   &50\%  & 0.002 & 1.000 & 1.012 & 0.957 \\ 
& 300 &0.70 & 0.65 & PL/wCI &50\% & 0.002 & 1.121 & 1.041 & 0.959 \\
& 300 &0.70 & 0.65 & PR/wCI &50\% &  0.002 & 1.126 & 1.047 & 0.959 \\ 
  \hline
  \hline
\end{tabular}
\end{table}

 \begin{table}[ht]
 \caption{Proportional hazards simulation results for the impact parameter equal to  0.10.
  The average bias and relative efficiency with respect to the projection-based CPE were averaged over 2,000 simulation iterations. Standard errors were calculated based on 50 bootstrap samples within each iteration and compared to the simulation standard error. Coverage was based on 95\% confidence intervals using the bootstrap standard error estimate.   }
\centering
\tiny
\begin{tabular}{lccccccccccccc} \label{PHsimresultsDelta010}
\vspace{-0.25cm}
   &&&&&&  & Relative Efficiency   & Estimated to  &\\
& N & Enhanced &  Projection & Estimate & Censoring &Bias & CPE w.r.t. wCI & Simulation SE & Coverage \\

 \hline
  \hline
  Projection &&&&&&&&& \\
& 300 &0.70 & 0.60 & PL/CPE  &0\% & 0.001 & 1.000 & 1.006 & 0.954 \\ 
& 300 &0.70 & 0.60 & PL/wCI &0\%  & 0.001 & 1.241 & 1.034 & 0.957 \\ 
& 300 &0.70 & 0.60 & PR/wCI &0\%  &   0.001 & 1.240 & 1.038 & 0.959 \\ 
\hline
& 300 &0.70 & 0.60 & PL/CPE  &25\% & 0.001 & 1.000 & 1.009 & 0.959 \\ 
& 300 &0.70 & 0.60 & PL/wCI &25\%  & 0.001 & 1.167 & 1.033 & 0.958 \\ 
& 300 &0.70 & 0.60 & PR/wCI &25\%  &  0.002 & 1.169 & 1.035 & 0.957 \\ 
\hline
& 300 &0.70 & 0.60 & PL/CPE  &50\%  &  0.002 & 1.000 & 0.996 & 0.948 \\ 
& 300 &0.70 & 0.60 & PL/wCI &50\%  &0.002 & 1.104 & 1.024 & 0.958 \\
& 300 &0.70 & 0.60 & PR/wCI &50\%  & 0.002 & 1.108 & 1.032 & 0.960 \\
  \hline
  \hline
  Impact &&&&&&&&& \\
&     300 &0.70 & 0.60 & PL/CPE   &0\%  &0.001 & 1.000 & 0.987 & 0.945 \\ 
& 300 &0.70 & 0.60 & PL/wCI &0\%  & 0.001 & 1.248 & 1.028 & 0.952 \\
& 300 &0.70 & 0.60 & PR/wCI &0\%  & 0.002 & 1.251 & 1.026 & 0.952 \\
\hline
&   300 &0.70 & 0.60 & PL/CPE   &25\%  & 0.001 & 1.000 & 1.000 & 0.953 \\ 
& 300 &0.70 & 0.60 & PL/wCI &25\%  &  0.001 & 1.196 & 1.016 & 0.951 \\ 
& 300 &0.70 & 0.60 & PR/wCI &25\%  & 0.002 & 1.201 & 1.014 & 0.950 \\ 
\hline
& 300 &0.70 & 0.60 & PL/CPE   &50\%  & 0.002 & 1.000 & 1.010 & 0.953 \\ 
& 300 &0.70 & 0.60 & PL/wCI &50\% &  0.001 & 1.109 & 1.044 & 0.955 \\ 
& 300 &0.70 & 0.60 & PR/wCI &50\% &0.002 & 1.110 & 1.051 & 0.958 \\ 
  \hline
  \hline
\end{tabular}
\end{table}

 \begin{table}[ht]
 \caption{Non-proportional hazards simulation results for the impact parameter equal to 0.025.
  The average bias and relative efficiency with respect to the projection-based CPE were averaged over 2,000 simulation iterations. Standard errors were calculated based on 50 bootstrap samples within each iteration and compared to the simulation standard error. Coverage was based on 95\% confidence intervals using the bootstrap standard error estimate. }
\centering
\tiny
\begin{tabular}{lccccccccccccc} \label{NPHsimresultsDelta0025}
\vspace{-0.25cm}
   &&&&&& & Relative Efficiency   & Estimated to  &\\
& N & Enhanced &  Projection & Estimate & Censoring &Bias & CPE w.r.t. wCI & Simulation SE & Coverage \\

 \hline
 \hline
  Projection &&&&&&&&& \\
&   300 &0.70 & 0.675 & PL/CPE & 0\% &  -0.107 & 1.000 & 0.737 & 0.006 \\ 
& 300 &0.70 & 0.675 & PL/wCI &0\%  &  -0.008 & 0.333 & 0.802 & 0.958 \\
& 300 &0.70 & 0.675 & PR/wCI &0\%  &  0.001 & 0.192 & 1.001 & 0.952 \\
\hline
& 300 &0.70 & 0.675 & PL/CPE  &25\% &  -0.053 & 1.000 & 0.959 & 0.270 \\ 
& 300 &0.70 & 0.675 & PL/wCI &25\%  &   0.001 & 0.377 & 1.009 & 0.942 \\ 
& 300 &0.70 & 0.675 & PR/wCI &25\%  &   0.002 & 0.376 & 0.999 & 0.940 \\
\hline
& 300 &0.70 & 0.675 & PL/CPE  &50\%  &   -0.016 & 1.000 & 0.965 & 0.875 \\ 
& 300 &0.70 & 0.675 & PL/wCI &50\%  &  0.002 & 0.794 & 1.004 & 0.946 \\ 
& 300 &0.70 & 0.675 & PR/wCI &50\%  &   0.002 & 0.795 & 1.003 & 0.945 \\
  \hline
  \hline
  Impact &&&&&&&&& \\
&  300 &0.70 & 0.675 & PL/CPE   &0\%  &  -0.013 & 1.000 & 1.010 & 0.607 \\ 
& 300 &0.70 & 0.675 & PL/wCI &0\%  &   -0.006 & 1.444 & 0.869 & 0.936 \\ 
& 300 &0.70 & 0.675 & PR/wCI &0\%  &  \ 0.002 & 0.741 & 1.077 & 0.959 \\ 
\hline
&   300 &0.70 & 0.675 & PL/CPE   &25\%  & -0.007 & 1.000 & 1.012 & 0.805 \\
& 300 &0.70 & 0.675 & PL/wCI &25\%  &  <0.001 & 0.984 & 1.091 & 0.945 \\
& 300 &0.70 & 0.675 & PR/wCI &25\%  & 0.002 & 0.981 & 1.066 & 0.949 \\ 
\hline
& 300 &0.70 & 0.675 & PL/CPE   & 50\%  &   -0.001 & 1.000 & 1.022 & 0.909 \\
& 300 &0.70 & 0.675 & PL/wCI &50\% &   0.001 & 1.033 & 1.065 & 0.946 \\
& 300 &0.70 & 0.675 & PR/wCI &50\% &   0.003 & 1.051 & 1.059 & 0.948 \\
  \hline
  \hline
\end{tabular}
\end{table}

 \begin{table}[ht]
 \caption{Non-proportional hazards simulation results for the impact parameter equal to 0.05.
  The average bias and relative efficiency with respect to the projection-based CPE were averaged over 2,000 simulation iterations. Standard errors were calculated based on 50 bootstrap samples within each iteration and compared to the simulation standard error. Coverage was based on 95\% confidence intervals using the bootstrap standard error estimate. }
\centering
\tiny
\begin{tabular}{lccccccccccccc} \label{NPHsimresultsDelta005}
\vspace{-0.25cm}
   &&&&&& & Relative Efficiency   & Estimated to  &\\
& N & Enhanced &  Projection & Estimate & Censoring &Bias & CPE w.r.t. wCI & Simulation SE & Coverage \\

 \hline
 \hline
  Projection &&&&&&&&& \\
&   300 &0.70 & 0.65 & PL/CPE & 0\% & -0.092 & 1.000 &   0.763 & 0.021 \\ 
& 300 &0.70 & 0.65 & PL/wCI &0\%  & -0.008 & 0.396 & 0.799 & 0.952 \\  
& 300 &0.70 & 0.65 & PR/wCI &0\%  & 0.001 & 0.227 & 1.007 & 0.952 \\ 
\hline
& 300 &0.70 & 0.65 & PL/CPE  &25\% &  -0.044 & 1.000 & 0.975 & 0.387 \\ 
& 300 &0.70 & 0.65 & PL/wCI &25\%  &   0.001 & 0.450 & 1.020 & 0.954 \\ 
& 300 &0.70 & 0.65 & PR/wCI &25\%  &   0.001 & 0.449 & 1.006 & 0.951 \\
\hline
& 300 &0.70 & 0.65 & PL/CPE  &50\%  & -0.012 & 1.000 & 0.987 & 0.913 \\ 
& 300 &0.70 & 0.65 & PL/wCI &50\%  & 0.001 & 0.875 & 1.015 & 0.953 \\
& 300 &0.70 & 0.65 & PR/wCI &50\%  &  0.002 & 0.877 & 1.013 & 0.954 \\  
  \hline
  \hline
  Impact &&&&&&&&& \\
&  300 &0.70 & 0.65 & PL/CPE   &0\%  & -0.031 & 1.000 &  0.934 & 0.254 \\ 
& 300 &0.70 & 0.65 & PL/wCI &0\%  & -0.006 & 0.809 & 0.868 & 0.938 \\ 
& 300 &0.70 & 0.65 & PR/wCI &0\%  & 0.002 & 0.470 & 1.058 & 0.959 \\ 
\hline
&   300 &0.70 & 0.65 & PL/CPE   &25\%  &  -0.016 & 1.000 & 1.002 & 0.772 \\ 
& 300 &0.70 & 0.65 & PL/wCI &25\%  & $<$0.001 & 0.746 & 1.079 & 0.963 \\
& 300 &0.70 & 0.65 & PR/wCI &25\%  & 0.002 & 0.746 & 1.059 & 0.959 \\ 
\hline
& 300 &0.70 & 0.65 & PL/CPE   & 50\%  &  -0.005 & 1.000 & 1.029 & 0.952 \\ 
& 300 &0.70 & 0.65 & PL/wCI &50\% & 0.001 & 0.984 & 1.064 & 0.964 \\ 
& 300 &0.70 & 0.65 & PR/wCI &50\% & 0.003 & 0.992 & 1.060 & 0.962 \\
  \hline
  \hline
\end{tabular}
\end{table}

\begin{table}[ht]
 \caption{Non-proportional hazards simulation results for the impact parameter equal to 0.10.
  The average bias and relative efficiency with respect to the projection-based CPE were averaged over 2,000 simulation iterations. Standard errors were calculated based on 50 bootstrap samples within each iteration and compared to the simulation standard error. Coverage was based on 95\% confidence intervals using the bootstrap standard error estimate.   }
\centering
\tiny
\begin{tabular}{lccccccccccccc} \label{NPHsimresultsDelta010}
\vspace{-0.25cm}
   &&&&&&  & Relative Efficiency   & Estimated to  &\\
& N & Enhanced &  Projection & Estimate & Censoring &Bias & CPE w.r.t. wCI & Simulation SE & Coverage \\
 \hline
 \hline
 Projection &&&&&&&&& \\
&  300 &0.70 & 0.65 & PL/CPE & 0\% &  -0.058 & 1.000 &0.861 & 0.146 \\ 
&300 &0.70 & 0.65 & PL/wCI &0\%  &-0.010 & 0.627 & 0.842 & 0.932 \\ 
&300 &0.70 & 0.65 & PR/wCI  &0\%  & 0.002 & 0.365 & 1.018 & 0.954 \\  
\hline
&300 &0.70 & 0.60 & PL/CPE  &25\% & -0.027 & 1.000 & 0.986 & 0.719 \\ 
&300 &0.70 & 0.60 & PL/wCI  &25\%  &  0.001 & 0.683 & 1.050 & 0.959 \\  
&300 &0.70 & 0.60 & PR/wCI &25\%  &   0.002 & 0.678 & 1.019 & 0.954 \\  
\hline
&300 &0.70 & 0.60 & PL/CPE  &50\%  &   -0.006 & 1.000 & 1.012 & 0.949 \\ 
&300 &0.70 & 0.60 & PL/wCI &50\%  & 0.002 & 0.986 & 1.039 & 0.962 \\
&300 &0.70 & 0.60 & PR/wCI &50\%  &   0.002 & 0.992 & 1.030 & 0.958 \\ 
  \hline
  \hline
  
  Impact &&&&&&&&& \\

&      300 &0.70 & 0.60 & PL/CPE   &0\%  & -0.066 & 1.000 &   0.855 & 0.071 \\ 
&300 &0.70 & 0.60 & PL/wCI &0\%  & -0.004 & 0.509 & 0.889 & 0.936 \\ 
&300 &0.70 & 0.60 & PR/wCI &0\%  & 0.002 & 0.320 & 1.047 & 0.959 \\
\hline
&  300 &0.70 & 0.60 & PL/CPE   &25\%  & -0.034 & 1.000 & 0.977 & 0.569 \\ 
&300 &0.70 & 0.60 & PL/wCI &25\%  &  $<$0.001 & 0.568 & 1.073 & 0.965 \\ 
&300 &0.70 & 0.60 & PR/wCI &25\%  &   0.002 & 0.562 & 1.045 & 0.959 \\
\hline
&300 &0.70 & 0.60 & PL/CPE   &50\%  & -0.012 & 1.000 & 1.016 & 0.928 \\ 
&300 &0.70 & 0.60 & PL/wCI &50\% & 0.001 & 0.901 & 1.057 & 0.962 \\
&300 &0.70 & 0.60 & PR/wCI &50\% & 0.003 & 0.901 & 1.052 & 0.957 \\ 
  \hline
  \hline
\end{tabular}
\end{table}

 \clearpage

\clearpage
\newpage

\beginsupplement

{\Large
\centering
\textbf{Supplementary Material} \\
Measuring the Impact of New Risk Factors Within Survival Models\\
}

\renewcommand{\theequation}{S.\arabic{equation}}  
\setcounter{equation}{0}   
\setcounter{page}{1}

\newpage
\clearpage

\section{ Assessment of the proportional hazards assumption in data example of 631 metastatic prostate cancer patients.}

 \begin{figure}[h]
\centering
 \includegraphics[width=5in, height=7.5in]{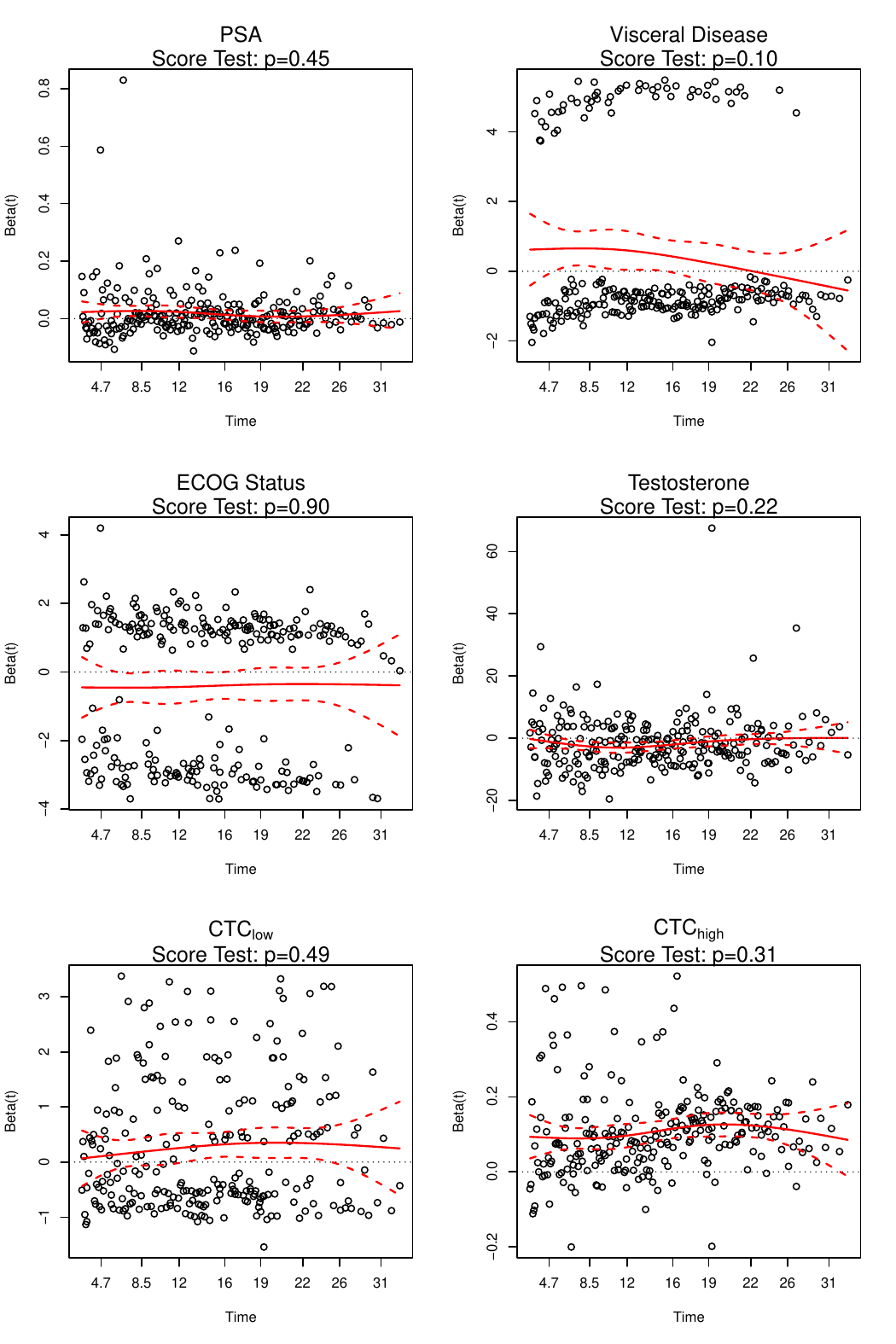} 
 \caption{\label{fig.phenhanced} The smoothed relationship between the scaled Schoenfeld residuals and time are plotted for each risk factor in the enhanced model. The p-value provided at the top of each plot is generated from a null test statistic of a constant slope. }
\end{figure}

 \begin{figure}[h]
\centering
 \includegraphics[width=5in, height=5in]{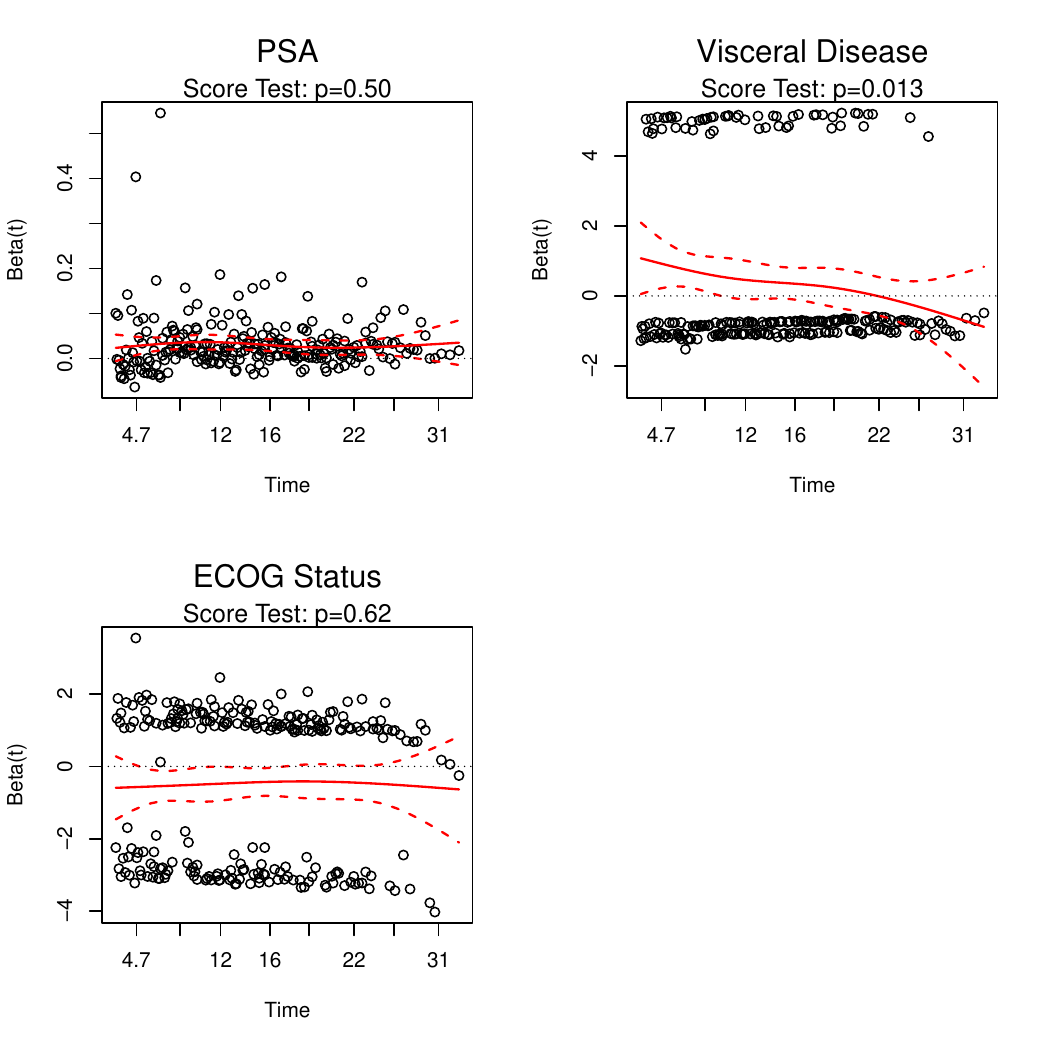} 
 \caption{\label{fig.phreduced}  The Schoenfeld residuals are plotted over time for each risk factor in the reduced model. The p-value provided at the top of each plot is generated from a null test statistic of a constant slope.}
\end{figure}  

\newpage
\clearpage

\section{Derivation of Theorems 1 and 2}

\noindent The following notation and regularity conditions are used throughout the appendix.

\vskip 0.20in

\noindent (C1)  The difference in risk indices are denoted by  

\noindent \hspace{0.30in} $\bb^T\bmx_{ij} = \bb^T\bmx_i - \bb^T\bmx_j$, $ \ \ \bg^T\bmz_{ij} = \bg^T\bmz_i - \bg^T\bmz_j$  

\noindent \hspace{0.30in} and the marginal precedence probability is represented as

\noindent \hspace{0.30in}  $\pi(\tau) = \mbox{Pr}(T_1 < T_2, T_1 < \tau)$.

\bigskip

\noindent (C2) $(T_i, \bmX_i, \bmZ_i), i = 1, \ldots, n$ are an independent random sample from $(T, \bmX, \bmZ)$.

\bigskip

\noindent (C3) The random vector $(\bmX, \bmZ)$ and the parameter vector $(\bb_0, \bg_0)$ each lie in a

\noindent \hspace{0.30in} $p+q$ dimensional bounded rectangle, where $(\bb_0, \bg_0)$ are the true values of $(\bb, \bg)$.

\bigskip

\noindent (C4) The enhanced model precedence probability $ \ \theta_{abcd}(\bb,\bg;\tau) = $
\small

\noindent \hspace{0.30in} $\mbox{Pr}(T_1 < T_2, T_1 < \tau |\bb^T\bmX_1=\bb^T\bmx_a, \bb^T\bmX_2=\bb^T\bmx_b,\bg^T\bmZ_1=\bg^T\bmz_c,\bg^T\bmZ_2=\bg^T\bmz_d) $
 
\normalsize 
\noindent \hspace{0.30in}  has bounded first partial derivatives in a compact neighborhood of $(\bb_0, \bg_0)$.

\bigskip

\noindent (C5)  As $n \rightarrow \infty$, the centered estimates

\noindent \hspace{0.50in} $\ n^{1/2} \left[(\hat{\bb} - \bb_0), (\hat{\bg} - \bg_0), (\hat{\theta}_{abcd}(\bb_0, \bg_0;\tau) - \theta_{abcd}(\bb_0, \bg_0;\tau)), (\hat{\pi}(\tau) - \pi_0(\tau) ) \right]$ 

\noindent \hspace{0.30in} converge to a multivariate normal random vector with mean zero.

\bigskip

\noindent (C6) $\phi(u/h)$ and  $\Phi(u/g)$ are a normal kernel density function with bandwidth $h$ and  

\noindent \hspace{0.30in}  a normal local distribution function with bandwidth $g$, respectively. 

\noindent \hspace{0.30in}  As $n \rightarrow \infty$, the bandwidths satisfy, 
 $h,g \rightarrow 0$, $ \ nh^2 \rightarrow \infty$, $ \ nh^4 \rightarrow 0$, 
 
 \noindent \hspace{0.30in}  and $ \ ng^4 \rightarrow 0$.


\vskip 0.50in

\noindent {\it Theorem 1}  
\[ \mbox{Let } \ \hat{\theta}_{12}^{[P]}(\hat{\bb};\tau) =  \frac{ n^{-2} \stackunder{k, l }{\sum} \hat{\theta}_{12kl}(\hat{\bb},\hat{\bg};\tau) \phi_h(\hat{\bb}^T\bmx_{1},\hat{\bb}^T\bmx_{k}) \phi_h(\hat{\bb}^T\bmx_{2},\hat{\bb}^T\bmx_{l}) }
 { n^{-2} \stackunder{k, l}{\sum} \phi_h(\hat{\bb}^T\bmx_{1},\hat{\bb}^T\bmx_{k}) \phi_h(\hat{\bb}^T\bmx_{2},\hat{\bb}^T\bmx_{l})}  \]  
 
\bigskip

\noindent \hspace{0.30in} Then $\ \hat{\theta}_{12}^{[P]}(\hat{\bb};\tau) = \mbox{Pr}(T_1 < T_2, T_1 < \tau |\bb_0^T \bmx_1, \bb_0^T \bmx_2) + o_p(1) $.

\vskip 0.25in

\noindent {\it Proof:}

\noindent The coefficient estimate $\hat{\bb}$ within the kernel $\phi(\cdot)$ may be treated as a constant since its $\sqrt{n}$ convergence rate is faster than the nonparametric kernel estimate rate.

\bigskip
 
\noindent The denominator of the Nadarya-Watson kernel estimate $\hat{\theta}^{[P]}_{12}$, may be represented asymptotically as
\[  n^{-2} \stackunder{k, l}{\sum} \phi_h(\hat{\bb}^T\bmx_{1},\hat{\bb}^T\bmx_{k}) \phi_h(\hat{\bb}^T\bmx_{2},\hat{\bb}^T\bmx_{l}) = f_{\beta_0^T X_1}(\bb_0^T \bmx_1)  f_{\beta_0^T X_2}(\bb_0^T \bmx_2) + o_p(1) .\]

\vskip 0.25in

\noindent For the numerator, a first order Taylor series expansion of the model precedence probability around $\left[\hat{\bb} = \bb_0, \ \hat{\bg} = \bg_0, \ \hat{\theta}_{abcd}(\bb_{0},\bg_{0};\tau ) = \theta_{abcd}(\bb_{0}, \bg_0;\tau )\right]$ results in

\bigskip  

\noindent \hspace{0.50in}$   n^{-2} \stackunder{k, l }{\sum} \hat{\theta}_{12kl}(\hat{\bb},\hat{\bg};\tau) \phi_h(\hat{\bb}^T\bmx_{1},\hat{\bb}^T\bmx_{k}) \phi_h(\hat{\bb}^T\bmx_{2},\hat{\bb}^T\bmx_{l}) $

\noindent \hspace{0.50in}$ = \mbox{Pr}(T_1 < T_2, T_1 < \tau |\bb_0^T \bmx_1, \bb_0^T \bmx_2) f_{\beta_0^T X_1}(\bb^T \bmx_1)  f_{\beta_0^T X_2}(\bb^T \bmx_2) + o_p(1) $.

\bigskip

\noindent Therefore, $ \ \ \hat{\theta}_{12}^{[P]}(\hat{\bb};\tau) = \mbox{Pr}(T_1 < T_2, T_1 < \tau |\bb_0^T \bmx_1, \bb_0^T \bmx_2) + o_p(1)$ .

\newpage

\noindent {\it Theorem 2:}

\bigskip

\noindent The following notation, used throughout the proof of Theorem 2, is collected here. 

\bigskip

\noindent Consider the concordance probability estimate (CPE) from the extended model and the CPE for $\bmX$ alone obtained by projection  
\[ K_n (\hat{\bb},\hat{\bg}, \hat{\theta}(\hat{\bb},\hat{\bg};\tau),\hat{\pi}(\tau)) = 
[n^{2}\hat{\pi}(\tau)]^{-1} \stackunder{i, j}{\sum} I(\hat{\bb}^T\bmx_{ij} + \hat{\bg}^T\bmz_{ij} > 0)  \ \hat{\theta}_{ijij}(\hat{\bb},\hat{\bg};\tau)  \]
\[ K_n^{[P]}(\hat{\bb}, \hat{\theta}^{[P]}(\hat{\bb};\tau),\hat{\pi}(\tau)) =  
[n^{2}\hat{\pi}(\tau)]^{-1} \stackunder{i, j}{\sum}  I(\hat{\bb}^T\bmx_{ij}>0) \hat{\theta}_{ij}^{[P]}(\hat{\bb};\tau) ,\]
where to make the arguments in the CPE statistics more concise, going forward we write
\[ \hat{\theta} \equiv \hat{\theta}(\hat{\bb},\hat{\bg};\tau) \ \ \ \ \ \hat{\theta}_0 \equiv \hat{\theta}(\bb_0, \bg_0;\tau)  \ \ \ \ \  \theta_0 \equiv \theta(\bb_0, \bg_0;\tau) \ \ \ \ \ \hat{\pi}(\tau) \equiv \hat{\pi} \ \ \ \ \ \pi_0(\tau) \equiv \pi_0  . \]

\noindent The impact parameter is denoted by 
\[ \xi(\tau) = \kappa(\tau) - \kappa^{[P]}(\tau) ,\]
where as described by equations (5) and (9) in the text, 
\[ K_n (\hat{\bb},\hat{\bg}, \hat{\theta},\hat{\pi};\tau) \ \stackrel{p}{\rightarrow} \ \kappa(\tau)  \] 
\[  K_n^{[P]} (\hat{\bb}, \hat{\theta}^{[P]},\hat{\pi};\tau) \ \stackrel{p}{\rightarrow} \ \kappa^{[P]}(\tau) . \]

\noindent Then Theorem 2 is stated as  
\begin{equation} n^{1/2} \left[K_n (\hat{\bb},\hat{\bg}, \hat{\theta},\hat{\pi};\tau) - K_n^{[P]}(\hat{\bb}, \hat{\theta}^{[P]},\hat{\pi};\tau) - \xi \left(\tau \right) \right]  \ \stackrel{D}{\rightarrow} \ N(0, V)  \end{equation}

\newpage

{\it Proof}

\noindent To derive this result, smooth versions of $K_n$ and $K_n^{[P]}$ are employed

\[ \tilde{K}_n(\hat{\bb},\hat{\bg}, \hat{\theta},\hat{\pi};\tau) = 
[n^2 \hat{\pi}(\tau)]^{-1} \stackunder{i, j }{\sum }\Phi \left(\frac{\hat{\bb}^T\bmx_{ij} + \hat{\bg}^T\bmz_{ij}}{g} \right)  \ \hat{\theta}_{ijij}(\hat{\bb},\hat{\bg};\tau)   \]
 
\[\tilde{K}_n^{[P]}(\hat{\bb},\hat{\theta}^{[P]},\hat{\pi};\tau) = [n^2 \hat{\pi}(\tau)]^{-1} \stackunder{i, j}{\sum}  \Phi \left(\frac{\hat{\bb}^T\bmx_{ij}}{g} \right) \ \hat{\theta}_{ij}^{[P]}(\hat{\bb};\tau)  \]

\vskip 0.25in

\noindent The asymptotic normal distribution (S.1) is demonstrated via the decomposition 

\small
\noindent $ n^{1/2} \left\{ \left[K_n(\hat{\bb},\hat{\bg}, \hat{\theta},\hat{\pi};\tau) - \tilde{K}_n(\hat{\bb},\hat{\bg}, \hat{\theta},\hat{\pi};\tau) \right] - 
\left[ K_n^{[P]}(\hat{\bb},\hat{\theta}^{[P]},\hat{\pi};\tau) - \tilde{K}_n^{[P]}(\hat{\bb},\hat{\theta}^{[P]},\hat{\pi};\tau) \right] \right\} + $

\noindent $n^{1/2} \left\{ \left[\tilde{K}_n(\hat{\bb},\hat{\bg}, \hat{\theta},\hat{\pi};\tau) - \tilde{K}_n(\hat{\bb},\hat{\bg}, \hat{\theta},\pi_0;\tau) \right] -
\left[\tilde{K}_n^{[P]}(\hat{\bb},\hat{\theta}^{[P]},\hat{\pi};\tau) - \tilde{K}_n^{[P]}(\hat{\bb},\hat{\theta}^{[P]},\pi_0;\tau) \right] \right\} + $

\noindent $n^{1/2} \left\{ \left[\tilde{K}_n(\hat{\bb},\hat{\bg}, \hat{\theta},\pi_0;\tau) - \tilde{K}_n(\bb_0,\bg_0, \hat{\theta}_0,\pi_0;\tau) \right] -
\left[\tilde{K}_n^{[P]}(\hat{\bb},\hat{\theta}^{[P]},\pi_0;\tau) - \tilde{K}_n^{[P]}(\bb_0, \hat{\theta}_0^{[P]},\pi_0;\tau) \right] \right\} + $

\noindent $ n^{1/2} \left\{ \left[\tilde{K}_n(\bb_0,\bg_0, \hat{\theta}_0,\pi_0;\tau) - \tilde{K}_n(\bb_0,\bg_0, \theta_0,\pi_0;\tau) \right] - 
\left[\tilde{K}_n^{[P]}(\bb_0,\hat{\theta}_0^{[P]},\pi_0;\tau) - \tilde{K}_n^{[P]}(\bb_0, \theta_0^{[P]},\pi_0;\tau) \right] \right\} + $

\noindent $n^{1/2} \left\{ \left[\tilde{K}_n(\bb_0,\bg_0, \theta_0,\pi_0;\tau) - \kappa(\tau) \right] - 
\left[ \tilde{K}_n^{[P]}(\bb_0,\theta_0,\pi_0;\tau) - \kappa^{[P]}( \tau) \right] \right\} $.

\normalsize

\noindent The asymptotic results are developed below.  

\vskip 0.25in

\noindent $\bullet$  From the first line of the decomposition, 

\noindent $n^{1/2} \left\{ \left[K_n(\hat{\bb},\hat{\bg}, \hat{\theta},\hat{\pi};\tau) - \tilde{K}_n(\hat{\bb},\hat{\bg}, \hat{\theta},\hat{\pi};\tau) \right] \right\} = $ 
\small
\[ n^{-3/2} [\hat{\pi}(\tau)]^{-1} \stackunder{i, j}{\sum} \left\{ I(\hat{\bb}^T\bmx_{ij} + \hat{\bg}^T\bmz_{ij} > 0) - \Phi \left(\frac{\hat{\bb}^T\bmx_{ij} + \hat{\bg}^T\bmz_{ij}}{g} \right) \right\} \hat{\theta}_{ijij}(\hat{\bb},\hat{\bg};\tau) .\]
\normalsize
Since $0 < \hat{\pi}(\tau) < 1$ and $\left| \hat{\theta}_{ijij}(\hat{\bb},\hat{\bg};\tau) \right| < 1$,
\[ < M n^{-3/2} \sup_{\beta, \gamma}  \left|\stackunder{i, j}{\sum} \left\{ I(\bb^T\bmx_{ij} + \bg^T\bmz_{ij} > 0) - \Phi \left(\frac{\bb^T\bmx_{ij} + \bg^T\bmz_{ij}}{g} \right) \right\} \right| \]
The right hand side of this inequality may be rewritten as
\[ M n^{1/2} \sup_{\beta, \gamma} \left| \int_u \left\{ I(u > 0) - \Phi \left(\frac{u}{g} \right) \right\} d\hat{F}_{n \times n}(u) \right| \]
where $\hat{F}_{n \times n}(u)$ is the empirical cumulative distribution function with jumps at each of the $n^2$ elements of $u_{ij}$. This expression has the same form as the expression in Heller (2007, Lemma A.1), where it is shown 
\[ n^{1/2} \left\{ \left[K_n(\hat{\bb},\hat{\bg}, \hat{\theta},\hat{\pi};\tau) - \tilde{K}_n(\hat{\bb},\hat{\bg}, \hat{\theta},\hat{\pi};\tau) \right] \right\}  \ \stackrel{p}{\rightarrow} \ 0  \]
uniformly in $(\bb, \bg)$.
	
\noindent The same argument demonstrates that	
\[ n^{1/2} \left\{ \left[K_n^{[P]}(\hat{\bb}, \hat{\theta}^{[P]},\hat{\pi};\tau) - \tilde{K}_n^{[P]}(\hat{\bb}, \hat{\theta}^{[P]},\hat{\pi};\tau) \right] \right\} \ \stackrel{p}{\rightarrow} \ 0 \]
uniformly in $(\bb, \bg)$.

\vskip 0.25in

\noindent $\bullet$   For the second line of the decomposition

\noindent $n^{1/2} \left\{ \left[\tilde{K}_n(\hat{\bb},\hat{\bg}, \hat{\theta},\hat{\pi};\tau) - \tilde{K}_n(\hat{\bb},\hat{\bg}, \hat{\theta},\pi_0;\tau) \right] \right\} =$
\[ \left\{ n^{1/2} \frac{[\pi_0(\tau)-\hat{\pi}(\tau)]}{\pi_0(\tau) \hat{\pi}(\tau)} \right\} \left\{ n^{-2} \stackunder{i, j }{\sum } \Phi \left(\frac{\hat{\bb}^T\bmx_{ij} + \hat{\bg}^T\bmz_{ij}}{g} \right)  \ \hat{\theta}_{ijij}(\hat{\bb},\hat{\bg};\tau) \right\} \]

\noindent The first term in curly brackets, from Kaplan-Meier estimation theory, converges in distribution to a mean zero normal random variable.  The second term in curly brackets converges in probability to a constant. It follows from Slutsky's Theorem that 
\[ n^{1/2} \left\{ \left[\tilde{K}_n(\hat{\bb},\hat{\bg}, \hat{\theta},\hat{\pi};\tau) - \tilde{K}_n(\hat{\bb},\hat{\bg}, \hat{\theta},\pi_0;\tau) \right] \right\}\ \stackrel{D}{\rightarrow} \ N(0, V_{2}) .\]

\noindent A similar argument gives
\[ n^{1/2} \left\{ \left[ K_n^{[P]}(\hat{\bb}, \hat{\theta}^{[P]},\hat{\pi};\tau) - \tilde{K}_n^{[P]}(\hat{\bb}, \hat{\theta}^{[P]},\pi_0;\tau) \right]  \right\} \ \stackrel{D}{\rightarrow} \ N(0, V_{2P}) .\]

\vskip 0.25in

\noindent $ \bullet$  In the third and fourth lines of the decomposition, using the asymptotic normality in (C5), a first order Taylor series around $\left[\hat{\bb}, \hat{\bg}, \hat{\theta}(\hat{\bb},\hat{\bg}) \right]  = \left[\bb_0, \bg_0, \theta_0 \right]$, results in 
\[n^{1/2} \left\{ \tilde{K}_n(\hat{\bb},\hat{\bg}, \hat{\theta},\pi_0;\tau) - \tilde{K}_n(\bb_0,\bg_0, \hat{\theta}_0,\pi_0;\tau)  \right\} \ \stackrel{D}{\rightarrow} \ N(0, V_{3})\]
\[n^{1/2} \left\{\tilde{K}_n^{[P]}(\hat{\bb}, \hat{\theta}^{[P]},\pi_0;\tau) - \tilde{K}_n^{[P]}(\bb_0, \hat{\theta}^{[P]}_0,\pi_0;\tau) \right\} \ \stackrel{D}{\rightarrow} \ N(0, V_{3P})\]
\[ n^{1/2} \left\{ \tilde{K}_n(\bb_0,\bg_0, \hat{\theta}_0,\pi_0;\tau) - \tilde{K}_n(\bb_0,\bg_0, \theta_0,\pi_0;\tau)  \right\} \ \stackrel{D}{\rightarrow} \ N(0, V_{4})\]
\[n^{1/2} \left\{ \tilde{K}_n^{[P]}(\bb_0, \hat{\theta}^{[P]}_0,\pi_0;\tau) - \tilde{K}_n^{[P]}(\bb_0,\bg_0, \theta^{[P]}_0,\pi_0;\tau) \right\} \ \stackrel{D}{\rightarrow} \ N(0, V_{4P})\]

\vskip 0.25in

\noindent $\bullet$ From the fifth line of the decomposition, 

\noindent $n^{1/2} \left[\tilde{K}_n(\bb_0,\bg_0, \theta_0,\pi_0;\tau) - \kappa(\tau) \right] $
\[ = n^{-3/2} \stackunder{i, j }{\sum} \left\{ [\pi_0(\tau)]^{-1} \left[\Phi \left(\frac{\bb^T\bmx_{ij} }{g} \right)  \ \theta_{ijij}(\bb_0,\bg_0;\tau) \right] - \kappa(\tau) \right\} \]
is a degree 2 U-statistic, and therefore
\[ n^{1/2} \left[\tilde{K}_n(\bb_0,\bg_0, \theta_0,\pi_0;\tau) - \kappa(\tau) \right]\ \stackrel{D}{\rightarrow} \ N(0, V_{5}) .\]   

\bigskip
\noindent For the projected difference

\noindent $n^{1/2}\left[ \tilde{K}_n^{[P]}(\bb_0, \theta^{[P]}_0,\pi_0;\tau) - \kappa^{[P]}(\tau) \right]  = $ \small
\[  n^{-3/2}  \stackunder{i, j }{\sum} \left[ \Phi \left(\frac{\bb_0^T\bmx_{ij} }{g} \right)  \left\{ \frac{  \stackunder{k, l }{\sum} \theta_{ijkl}(\bb_0,\bg_0; \tau) \phi_h(\bb_0^T\bmx_{i},\bb_0^T\bmx_{k}) \phi_h(\bb_0^T\bmx_{j},\bb_0^T\bmx_{l}) }
 { \pi_0(\tau) \stackunder{k, l}{\sum} \phi_h(\bb_0^T\bmx_{i},\bb_0^T\bmx_{k}) \phi_h(\bb_0^T\bmx_{j},\bb_0^T\bmx_{l})} \right\}  - \kappa^{[P]}(\tau)\right] .\] \normalsize

\bigskip

\noindent \hspace{0.50in} Let $ \ a_{1ij}(\bb_0;\tau) = n^{-2} \stackunder{k, l}{\sum} \theta_{ijkl}(\bb_0,\bg_0;\tau) \phi_h(\bb_0^T\bmx_{i},\bb_0^T\bmx_{k}) \phi_h(\bb_0^T\bmx_{j},\bb_0^T\bmx_{l})$

\noindent \hspace{0.85in}  $a_{0ij}(\bb_0;\tau) = n^{-2} \stackunder{k, l}{\sum} \phi_h(\bb_0^T\bmx_{i},\bb_0^T\bmx_{k}) \phi_h(\bb_0^T\bmx_{j},\bb_0^T\bmx_{l}) $

\noindent \hspace{0.85in}  $\alpha_{mij}(\bb_0;\tau) = \lim_{n \rightarrow \infty} a_{mij}(\bb_0;\tau) \ \ \ \ \ \ \ \ m=0,1$

\bigskip

\noindent Then 
\[ n^{1/2}\left[ \tilde{K}_n^{[P]}(\bb_0, \theta^{[P]}_0,\pi_0;\tau) - \kappa^{[P]}(\tau) \right] = 
n^{-3/2} \stackunder{i, j }{\sum} \left[ \frac{\Phi \left(\frac{\bb_0^T\bmx_{ij} }{g} \right) a_{1ij}(\bb_0;\tau)}{\pi_0(\tau)  a_{0ij}(\bb_0;\tau)} - \kappa^{[P]}(\tau) \right] \]

\vskip 0.25in
\noindent Taylor expanding the ratio $a_{1ij}(\bb_0;\tau)/a_{0ij}(\bb_0;\tau)$ around $\alpha_{1ij}(\bb_0;\tau)/\alpha_{0ij}(\bb_0;\tau)$,

\bigskip

\small 
\noindent $n^{1/2}\left[ \tilde{K}_n^{[P]}(\bb_0, \theta^{[P]}_0,\pi;\tau) - \kappa^{[P]}(\tau) \right] = n^{-7/2}  \ \times $ \footnotesize
\[  \stackunder{i, j, k ,l }{\sum}  \left(\Phi \left(\frac{\bb_0^T\bmx_{ij} }{g} \right) \left\{ \frac{\alpha_{1ij}(\bb_0;\tau)}{\pi_0(\tau) \alpha_{0ij}(\bb_0;\tau)} + \frac{\phi_h(\bb_0^T\bmx_{i},\bb_0^T\bmx_{k}) \phi_h(\bb_0^T\bmx_{j},\bb_0^T\bmx_{l})}{\pi_0(\tau) \alpha_{0ij}(\bb_0;\tau)} \left[\theta_{ijkl}(\bb_0,\bg_0;\tau) - \frac{\alpha_{1ij}(\bb_0;\tau)}{\alpha_{0ij}(\bb_0;\tau)}\right] \right\} - \kappa^{[P]}(\tau) \right) \]

\normalsize
which is a degree 4 U-statistic and therefore
\[ n^{1/2}[\tilde{K}_n^{[P]}(\bb_0, \theta^{[P]}_0,\pi;\tau) - \kappa^{[P]}(\tau)] \ \stackrel{D}{\rightarrow} \ N(0, V_{5P}) .\]

\bigskip

\noindent Therefore, combining the terms of the decomposition,
\[ n^{1/2} \left[K_n (\hat{\bb},\hat{\bg}, \hat{\theta},\hat{\pi};\tau) - K_n^{[P]}(\hat{\bb}, \hat{\theta}^{[P]},\hat{\pi};\tau) - \xi \left(\tau \right) \right]  \ \stackrel{D}{\rightarrow} \ N(0, V) .\]

\newpage

	 \section{Assessment of the proportional hazards assumption in the simulation scenarios}
 
 \begin{table}[ht]
 \caption{The proportion of times the proportional hazards test proposed by \cite{Grambsch1994} for the enhanced model was rejected at the 0.05 level across all simulation scenarios. The proportion is calculated over the 2,000 simulation iterations in each scenario. }
\centering
\small
\begin{tabular}{lcccccccccccc} \label{simPHtest}
 \vspace{-0.25cm}
 &&&& PH Test& \\
Simulation Setting & Enhanced &  Projection  & Censoring & Proportion $P < 0.05$\\
 \hline
  \hline
 Proportional Hazards &0.70 & 0.675  & 0\% & 0.050 \\ 
Proportional Hazards  &0.70 & 0.675  &25\% &  0.056 \\ 
Proportional Hazards  &0.70 & 0.675  &50\%  & 0.056 \\ 
  \hline
Proportional Hazards &0.70 & 0.65  & 0\% & 0.065 \\ 
Proportional Hazards  &0.70 & 0.65  &25\% & 0.055 \\ 
Proportional Hazards  &0.70 & 0.65  &50\%  & 0.053 \\ 
 \hline
Proportional Hazards  &0.70 & 0.60  & 0\% & 0.067 \\ 
Proportional Hazards  &0.70 & 0.60   &25\% & 0.053  \\ 
Proportional Hazards  &0.70 & 0.60   &50\%  & 0.057 \\ 
  \hline
     \hline
Non-proportional Hazards &0.70 & 0.675  & 0\% & 1.00 \\ 
Non-proportional Hazards  &0.70 & 0.675  &25\% & 0.998  \\ 
Non-proportional Hazards  &0.70 & 0.675 &50\%  & 0.922   \\ 
   \hline
Non-proportional Hazards &0.70 & 0.65  & 0\% & 1.00 \\ 
Non-proportional Hazards  &0.70 & 0.65  &25\% & 0.998  \\ 
Non-proportional Hazards  &0.70 & 0.65 &50\%  &  0.933 \\ 
 \hline
Non-proportional Hazards  &0.70 & 0.60  & 0\% & 1.00\\ 
Non-proportional Hazards  &0.70 & 0.60   &25\% &  0.999\\ 
Non-proportional Hazards  &0.70 & 0.60   &50\%  & 0.927\\ 
  \hline
  \hline
\end{tabular}
\end{table}

\end{document}